\documentclass[graybox]{svmult}
\pdfoutput=1

\usepackage{mathptmx}       
\usepackage{helvet}         
\usepackage{courier}        
\usepackage{type1cm}        
\usepackage{makeidx}         
\usepackage{graphicx}        
\usepackage{multicol}        
\usepackage[bottom]{footmisc}
\usepackage{hyperref}        
\usepackage{soul}            
\hypersetup{colorlinks=true,urlcolor=blue}
\usepackage[square,numbers]{natbib}
\makeindex             

\begin{document}
\title*{Pair Production Detectors for Gamma-ray Astrophysics}
\author{David J. Thompson \thanks{corresponding author} and Alexander A. Moiseev}
\institute{David J. Thompson \at NASA Goddard Space Flight Center, Greenbelt, MD 20771  USA, \email{david.j.thompson@nasa.gov}
\and Alexander A. Moiseev \at University of Maryland, College Park, MD, USA \email{amoiseev@umd.edu}}
%
%
\maketitle
\abstract{}

Electron-positron pair production is the essential process for high-energy $\gamma$-ray astrophysical observations. Following the pioneering {\it OSO-3} counter 
telescope, the field evolved into use of particle tracking instruments, largely derived from high-energy physics detectors.  Although many of the techniques were developed on balloon-borne $\gamma$-ray telescopes, the need to escape the high background in the atmosphere meant that the breakthrough discoveries came from the {\it SAS-2} and {\it COS-B} satellites. The next major pair production success was EGRET on the {\it Compton Gamma Ray Observatory}, which provided the first all-sky map at energies above 100 MeV and found a variety of $\gamma$-ray sources, many of which were variable.  The current generation of pair production telescopes, {\it AGILE} and {\it Fermi} LAT, have broadened high-energy $\gamma$-ray astrophysics with particular emphasis on multiwavelength and multimessenger studies.  A variety of options remain open for future missions based on pair production with improved instrumental performance.  

\section*{Keywords} 
Gamma rays; High-energy astrophysics; Active galactic nuclei; Pulsars; Gamma-ray bursts; Gamma-ray telescopes; Pair production

\section{Introduction}
\label{Introduction}

For energies above a few tens of MeV, photons interact with matter, other photons, or magnetic fields primarily  by electron-positron pair production ($\gamma \rightarrow e^- + e^+$). Pair production is an explicit illustration of Einstein's $E = mc^2$, where the energy $E$ of the photon is converted into two particles with mass m, with the speed of light c being the conversion factor. This physical process has important implications for detection of astrophysical $\gamma$ rays:

\begin{enumerate}
    \item High-energy (greater than about 50 MeV) $\gamma$ rays cannot be reflected or refracted; there are no mirrors or lenses for these photons. 
    \item Gamma rays coming from space interact in Earth's upper atmosphere; therefore direct detection can only be done from space or from the edge of the atmosphere. 
    \item Properties of high-energy $\gamma$ rays can only be derived from measurements of the electron and positron resulting from pair production.  Gamma-ray instrumentation at these energies consists of charged particle detectors. 
\end{enumerate}

These considerations drive the most important factor in design of high-energy $\gamma$-ray telescopes. {\bf The real challenge is not in detecting the electron-positron pair. The critical issue is separating the cosmic pair-production events from background.} Thanks largely to developments in particle physics, various types of highly efficient particle detectors are available, and many of these have been applied to astrophysical $\gamma$-ray instruments.  Background comes in two forms:
 
\begin{enumerate}
    \item Space is filled with charged particles. Cosmic rays, solar energetic particles, and trapped particles in Earth's magnetic field outnumber high-energy $\gamma$ rays by orders of magnitude.  These charged particles are highly penetrating for most energies in the pair production regime, so shielding is impractical.  Such particles can masquerade as $\gamma$-ray interaction products.
    \item Many of the charged particles in space have enough energy to undergo inelastic nuclear interactions, producing secondaries such as neutral pions, which decay very quickly into $\gamma$ rays in the same energy range as seen by pair-production detectors. These secondary $\gamma$ rays are indistinguishable from cosmic $\gamma$ rays. The target material for such interactions can be local, as part of the detector or supporting structure, or it can be more diffuse, such as Earth's atmosphere. 
\end{enumerate}

Building detectors to deal with these issues was initially stimulated by the recognition that charged-particle cosmic rays in the Galaxy must interact with the interstellar gas to produce $\gamma$ rays in the MeV to GeV energy range \cite{Hayakawa,Morrison}. Since that time, detector technologies and access to space have undergone dramatic changes, and pair production telescopes have revealed a broad array of high-energy $\gamma$-ray sources in addition to the diffuse Galactic radiation that provided the original impetus for the field.  In this chapter, we review various approaches that have been taken or proposed as ways to use pair production to conduct astrophysical research. Section \ref{Counters} describes the early counter instruments; sections \ref{First_Generation} and \ref{Second_Generation} present the early imaging $\gamma$-ray telescopes; section \ref{Solid_State} outlines the current state of instrumentation; and section \ref{Future} covers some aspects of the future of pair-production telescopes. Additional information can be found in earlier review articles\cite{Kanbach,Tavani_review}. 
\section{Counter Detectors}
\label{Counters}

\begin{figure}[t]
    \centering
    \includegraphics[width=0.85\textwidth]{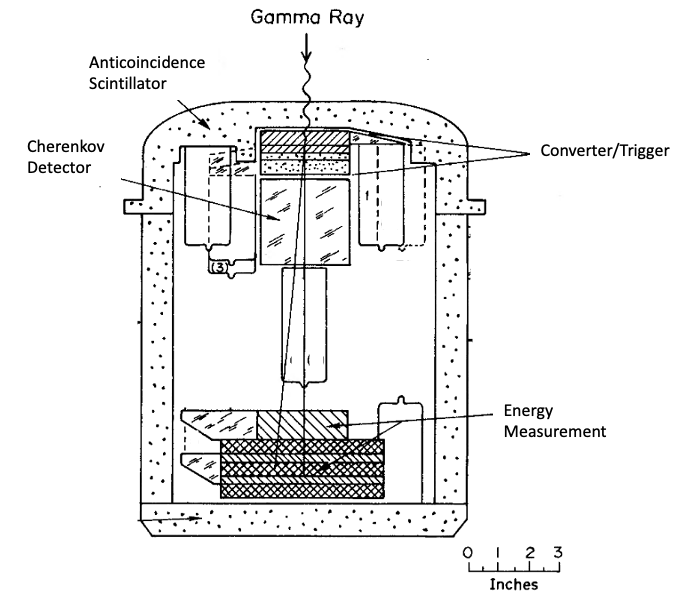}
    \caption{\small Simplified schematic diagram of the OSO-3 $\gamma$-ray instrument, adapted from \cite{OSO-3}.   }
    \label{fig:OSO-3}
\end{figure}

The 1960s saw the emergence of several types of pair-production telescopes, many of which were carried on high-altitude research balloons.  The greatest success from this decade was the use of scintillation counters on satellites.  The first hints of cosmic $\gamma$-ray detection using this method came from {\it Explorer XI} \cite{explorer}, and the real breakthrough came from a counter $\gamma$-ray instrument on the {\it OSO-3} satellite \cite{OSO-3}. 

Figure \ref{fig:OSO-3} is a scale drawing of the {\it OSO-3} $\gamma$-ray telescope, showing the variety of instrumentation used to detect the photons and reject the background. The plastic scintillator read out by photomultipliers and surrounding the detectors provides a first-level rejection of charged particles.  Plastic scintillator is highly efficient in particle detection while offering minimal absorption of $\gamma$ rays.  A $\gamma$ ray entering through the top of the instrument undergoes pair production in a Cesium Iodide (CsI) or plastic scintillator layer.  The electron-positron pair produce a signal in the directional Lucite Cerenkov counter.  The directionality discriminates against upward-moving particles.  Finally, the particles deposit energy in the layers of tungsten and Sodium Iodide (NaI), allowing a measurement of the original $\gamma$ ray energy. An experimental calibration of the instrument showed an effective energy threshold of about 50 MeV and a peak effective area of about 9 cm$^2$, with an angular response having a Full Width Half Maximum of about 24$^{\circ}$. No arrival direction information for individual photons within that wide opening angle was possible.  

In an era when rocket reliability was uncertain, three copies of this telescope were made.  The first was lost when its satellite failed to reach orbit.  The second was the one on {\it OSO-3}, launched into a low-Earth orbit on a Delta rocket in 1967 March.  The third was used for calibration.  The {\it OSO} satellites were spin stabilized, with the $\gamma$-ray instrument located in the rotating part of the satellite.  As the satellite precessed to follow the Sun (its primary mission), the rotation allowed the $\gamma$-ray telescope to sweep out the entire sky over the 16 months of operation, ending when the onboard tape recorders failed. 

During the mission, 621 $\gamma$ rays from the sky and a much larger number of atmospheric $\gamma$ rays were detected. Three important results emerged:

\begin{itemize}
\item As expected, atmospheric $\gamma$ rays vastly outnumber cosmic photons in the energy range above 50 MeV.  The Earth limb is particularly bright, and the east-west effect from geomagnetic screening of the positively charged cosmic rays is visible. 
\item The sky events show a clear peak toward the Galactic equator and a concentration toward the central region of the Milky Way, confirming the idea of high-energy $\gamma$ rays being produced by cosmic-ray interactions with interstellar gas. 
\item An apparently isotropic emission is visible, with $\gamma$ rays arriving from all directions in the sky.
\end{itemize}

The {\it OSO-3} results represented a milestone.  These were the first very-high-confidence observations in high-energy $\gamma$-ray astrophysics. Although the statistics were limited and the angular response quite broad, the careful design of the instrument demonstrated that the challenge of measuring cosmic $\gamma$ rays in a high-background environment could be met.

\section{First Generation Imaging Detectors}
\label{First_Generation}

Counter $\gamma$-ray detectors do not take full advantage of the information contained in the electron-positron pair resulting from $\gamma$-ray interactions.  Imaging of the individual pair-production events provides two additional important pieces of information:

\begin{enumerate}
    \item Visualizing the particle pair originating in a detector offers a distinctive signature of $\gamma$-ray pair production. The inverted ``V'' or ``Y'' pattern is extremely unlikely to originate from a single charged particle, thus providing an additional valuable discriminator against the huge charged-particle background in the space environment.  
    \item Measurement of the electron-positron pair provides information about the arrival direction of the incident $\gamma$ ray. The created particles retain information about both the energy and momentum of the original photon. 
\end{enumerate}

Deriving the properties of the incoming photon from measurements of the electron and positron involves inherent uncertainties and tradeoffs.  Due to conservation of energy and momentum, pair production cannot occur in free space.  Some momentum is lost to (usually) an atomic nucleus in the initial interaction. The particles also suffer collisional and radiative losses in passing through whatever target material comprises the detector. Thicker detectors offer higher probability for a photon to undergo pair production but more loss of directional information, primarily due to multiple Coulomb scattering of the electron and positron.  The balance between detection efficiency and directional information is a key element in astrophysical pair-production instrument design.

\subsection{Pioneering Balloon Instruments}

The first efforts to detect high-energy  cosmic $\gamma$ rays through pair production were small instruments carried on high-altitude research balloons. Such balloons can carry instruments to altitudes where the residual atmosphere above the balloon is only a few g cm$^{-2}$.  This is sufficiently little material that $\gamma$ rays arriving from outside the atmosphere have a relatively small probability of interacting before reaching the detector.  Although balloon flights have relatively short durations (days), they offer access to the near-space environment at a small fraction of the cost of a satellite. Particularly for an unproven field of research like high-energy  $\gamma$-ray astrophysics, balloons offered an opportunity to test pioneering techniques. 

\begin{figure}[t]
    \centering
    \includegraphics[width=0.7\textwidth]{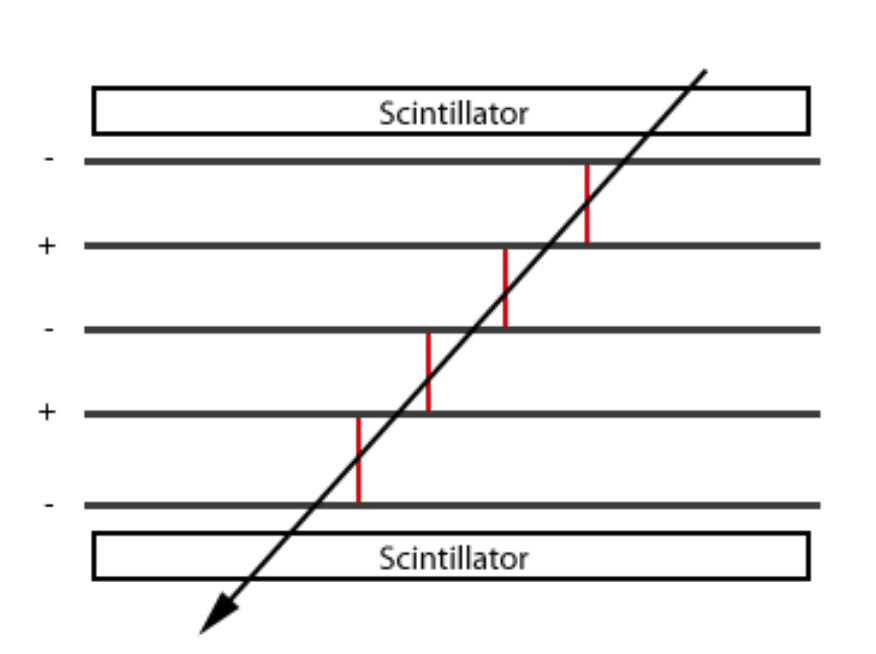}
    \caption{\small Schematic of the spark chamber principle. Reproduced with permission\cite{Collins}. }
    \label{fig:spark_chamber}
\end{figure}

In the 1960's, particle physicists' track imaging detector of choice was the spark chamber. Unlike bubble chambers, cloud chambers, or nuclear emulsions, spark chambers can be triggered by an external signal, making them useful for operation in a high-background environment. Figure \ref{fig:spark_chamber} illustrates the basic principle.  A series of conducting layers is stacked in a volume filled with a noble gas (typically neon/argon).  When a charged particle passes through the detector, it ionizes the gas, while at the same time producing signals in a triggering circuit, in this case a pair of scintillators.  The trigger then applies a high voltage to alternating layers, while grounding the ones between.  Sparks follow the ionization path, and  measurement of the spark locations tracks the particle. 

A number of groups pursuing high-energy  $\gamma$-ray astrophysics realized that the spark chamber enabled construction of pair-production instruments that could visualize the electron-positron tracks.  These first-generation imaging detectors employed various triggering systems and readout methods. 

\begin{figure}[t]
    \centering
    \includegraphics[width=0.8\textwidth]{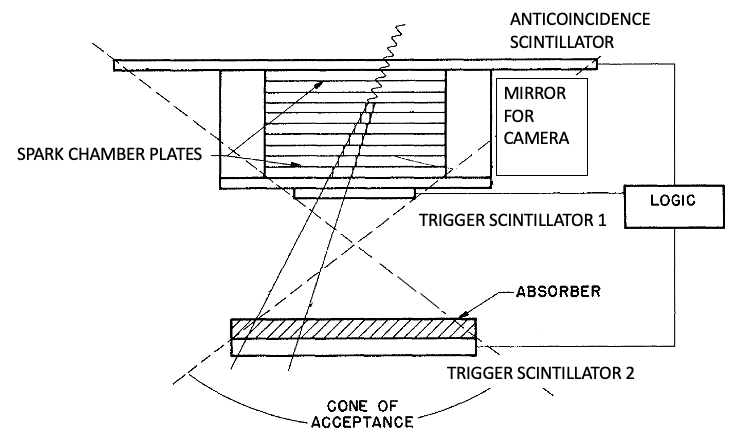}
    \caption{\small Schematic of an early spark chamber $\gamma$-ray telescope, adapted from\cite{Frye_Smith}.  The trigger logic was no signal in the anticoincidence scintillator plus coincident signals in the two trigger scintillators. }
    \label{fig:Frye_Smith}
\end{figure}

One of the early balloon-borne spark chamber instruments is shown schematically in Fig. \ref{fig:Frye_Smith}\cite{Frye_Smith}. The incoming $\gamma$-ray produces no signal in the anticoincidence scintillator, then undergoes pair production in one of the steel plates of the spark chamber.  The electron and positron produce a coincident signal in trigger scintillators 1 and 2, triggering the high voltage to produce the sparks. Stereoscopic views of the sparks were recorded on film, to be analyzed by manual inspection.  The instrument could measure $\gamma$ rays arriving up to 30$^{\circ}$ from the vertical, with the arrival directions of 100 MeV photons accurate to 3$^{\circ}$. This instrument flew twice in 1964 from the National Center for Atmospheric Research balloon base in Palestine, Texas (later the National Scientific Balloon Facility and now NASA's Columbia Scientific Balloon Facility).  As for many of the early balloon instruments, the flights produced only upper limits for $\gamma$ rays coming from sources outside the atmosphere. 

A later version of this instrument incorporated two features to help reduce background: scintillators were added to the sides of the spark chamber, and the lower simple counter was replaced by a directional Cherenkov counter (to discriminate against upward-moving particles)\cite{Frye_Wang}. A still-later version included some thicker plates to increase the conversion efficiency and some thinner plates to lower the effective energy threshold\cite{Albats}.  This instrument produced evidence of pulsed high-energy  $\gamma$ rays from the Crab pulsar.

\begin{figure}[t]
    \centering
    \includegraphics[width=0.7\textwidth]{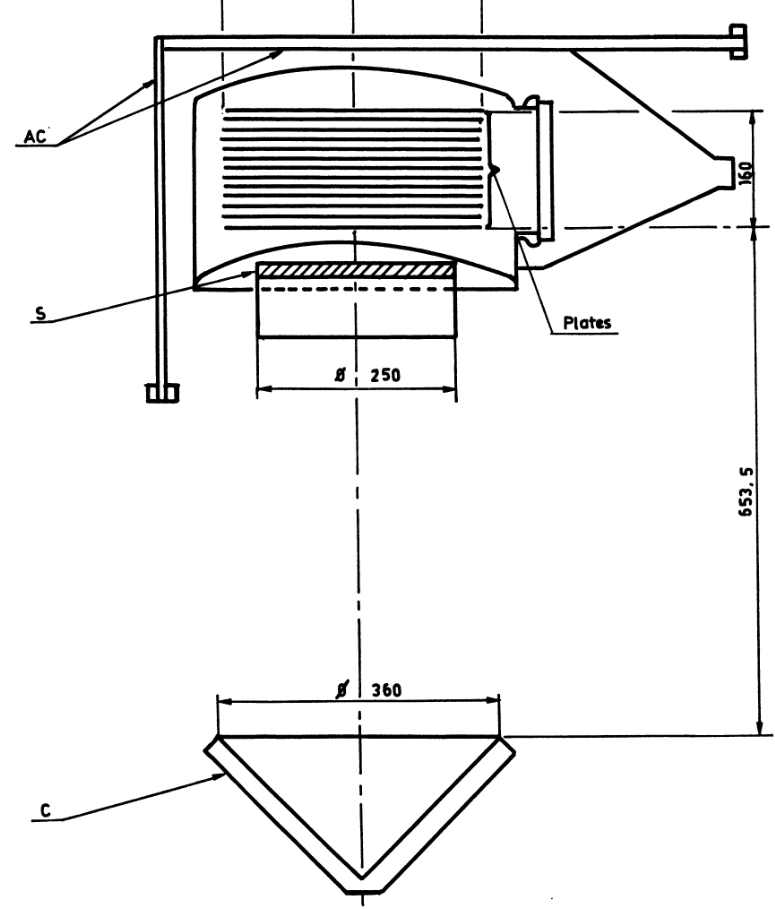}
    \caption{\small Schematic of another spark chamber $\gamma$-ray telescope, adapted from \cite{Leray}.  The trigger logic was no signal in the anticoincidence detector AC plus coincident signals in a scintillator S and Cherenkov detector C. Reproduced with permission \textcopyright ESO. }
    \label{fig:Leray}
\end{figure}

The earliest indications of pulsed high-energy  Crab $\gamma$-ray emission came from another balloon instrument with similar design\cite{Browning}, also flown from Palestine, Texas. It also had an optical spark chamber with film data recording and a three-element trigger: anticoincidence detectors surrounding five sides of the detector plus coincidence between a scintillator and a directional Cherenkov counter. A similar instrument flown on balloons in 1972, 1974, and 1976 obtained a signal from the high-mass X-ray binary Cygnus X-3 at energies above 40 MeV\cite{Galper_early} from one of the flights.

Still another instrument using similar technology is shown in Fig. \ref{fig:Leray}\cite{Leray}. The spark chamber was optical with film recording. The trigger was a twofold coincidence (mid-instrument scintillator and directional Cherenkov) with anticoincidence scintillators on top and three sides. Based on six successful balloon flights in 1969 from southern France, this instrument yielded a hint of signal from the Crab Pulsar, but not at a statistically significant level.

\begin{figure}[t]
    \centering
    \includegraphics[width=0.7\textwidth]{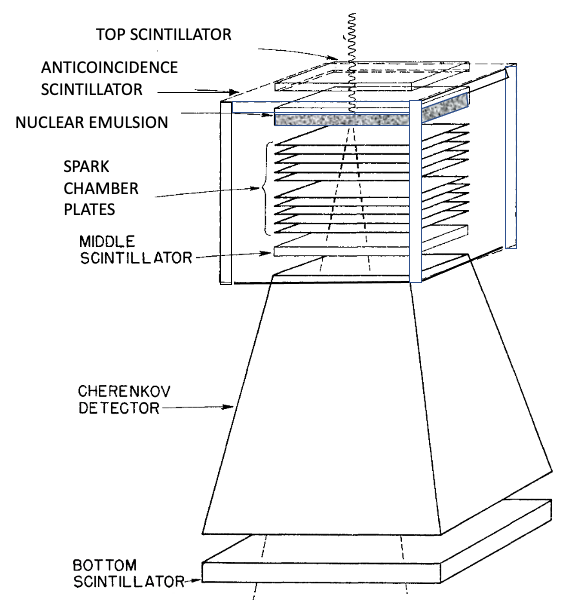}
    \caption{\small Schematic of an emulsion plus spark chamber $\gamma$-ray telescope, adapted from \cite{May}. }
    \label{fig:May}
\end{figure}

\begin{figure}[t]
    \centering
    \includegraphics[width=0.7\textwidth]{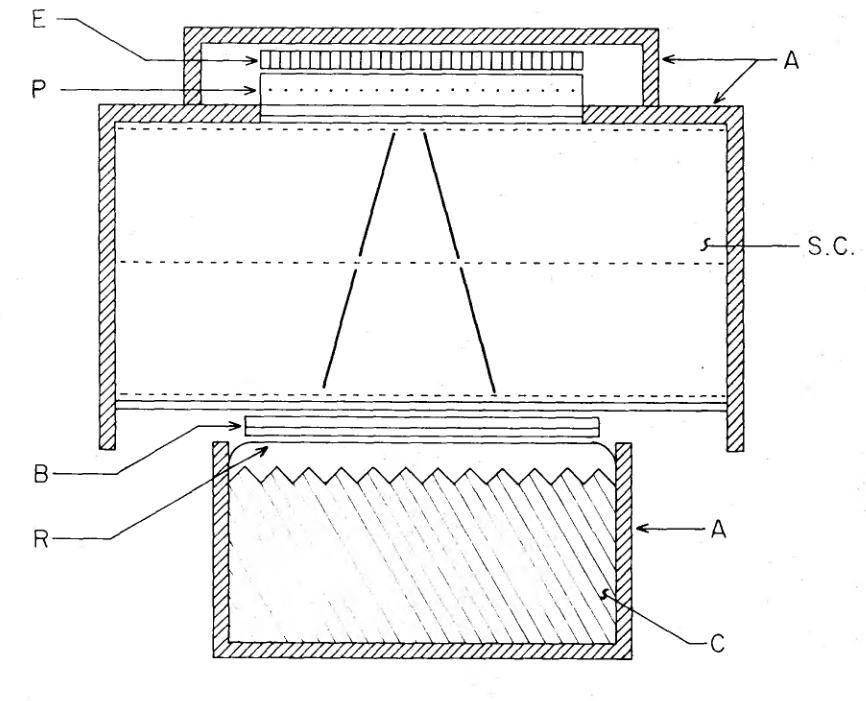}
    \caption{\small Schematic of another emulsion plus spark chamber $\gamma$-ray telescope, in this case using a wide-gap spark chamber instead of a stack. A: anticoincidence scintillator; E: emulsion; P: multiwire proportional chamber; S.C.: wide-gap spark chamber; B: trigger scintillators (2); C: Cherenkov detector; R: reflector for Cherenkov light \textcopyright AAS. Reproduced with permission\cite{NRL}. }
    \label{fig:NRL}
\end{figure}

One characteristic of all these detectors just described was that the pair production interaction took place in metal plates.  The scattering of the electron and positron limited the angular resolution of the measurements. This effect was greatest for energies below 100 MeV, where the scattering significantly changed the particle directions.  An approach to avoid this constraint was to use a nuclear emulsion as the converter material.  Fine-grained emulsions allowed measurement of the $\gamma$-ray arrival direction before significant scattering of the electron-positron pair. 

Two examples of detectors using emulsions are shown in Fig. \ref{fig:May}\cite{May} and \ref{fig:NRL}\cite{NRL}. The triggering was similar to the previous instruments: anticoincidence scintillator to discriminate against charged particles plus a scintillator/Cherenkov detector coincidence to trigger the spark chamber. The tracks in the spark chamber were then used to point back to the part of the emulsion where the pair-production event occurred.  Scanning of the emulsion could then resolve the incident direction of the $\gamma$ ray more accurately than a measurement using only a spark chamber. A balloon flight from Paran\'a, Argentina, in 1971 confirmed the {\it OSO-3} $\gamma$-ray excess along the Galactic plane surrounding the Galactic Center, at  energy down to 15 MeV\cite{NRL}.

All these balloon-borne instruments that used film or emulsions required recovery of the instrument before any data analysis could be done.  They also depended on extensive manual scanning to convert the images into quantifiable information about the electron-positron pair.  While suitable for balloon flights with durations of hours, these approaches would not be applicable to satellite instruments. What was needed was a way to digitize the information on board in such a way that the data could be telemetered to the ground. 

\begin{figure}[t]
    \centering
    \includegraphics[width=0.7\textwidth]{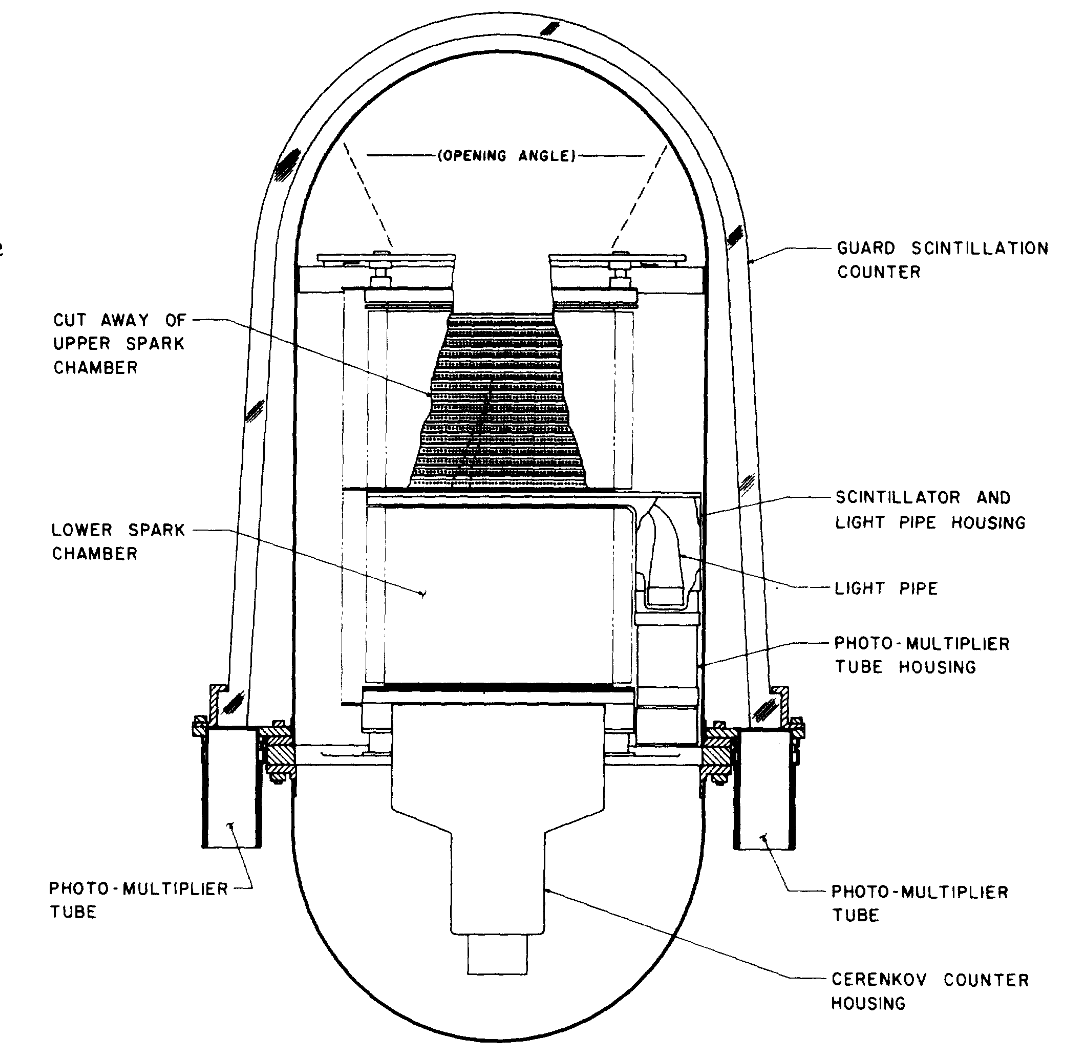}
    \caption{\small Schematic of an early wire grid spark chamber $\gamma$-ray telescope\cite{GSFC_6}. \textcopyright AAS. Reproduced with permission. }
    \label{fig:GSFC_6}
\end{figure}

The first spark chamber pair-production instrument designed explicitly as a prototype of a satellite telescope (Fig. \ref{fig:GSFC_6}\cite{GSFC_6}) had several features that differed from most of the other early designs:

\begin{itemize}
\item The spark chamber modules themselves were wire grids instead of plates. Each module had orthogonal planes of wires separated by about 2 mm.  Each wire threaded a magnetic core.  When the spark occurred, the set cores would record $x$ and $y$ coordinates that could be transmitted to the ground. 
\item The anticoincidence scintillator was a monolithic dome instead of tiles, providing  complete coverage of the active detector.
\item The pair production took place in separate thin plates made of gold plated onto aluminum for support. 
\item The middle triggering scintillator was located within the spark chamber stack instead of coming at the bottom, enabling measurement of the tracks before and after this trigger element. 
\end{itemize}

\begin{figure}[t]
    \centering
    \includegraphics[width=0.85\textwidth]{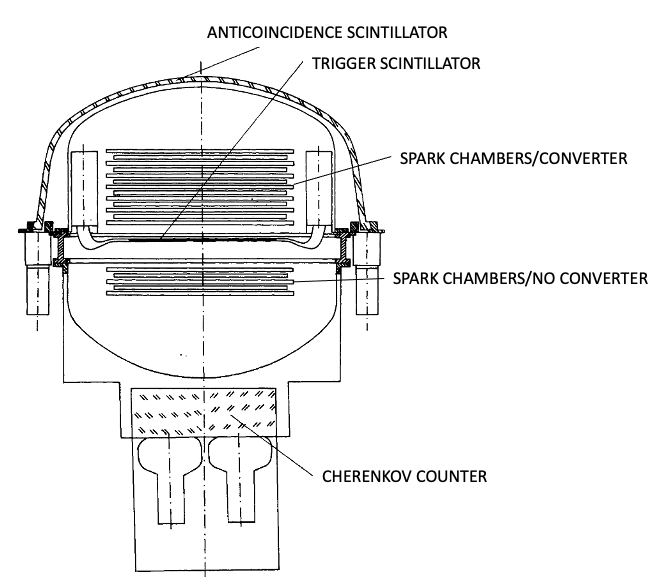}
    \caption{\small Schematic of a wire grid spark chamber $\gamma$-ray telescope  \cite{MPE}. The trigger was no signal in the anticoincidence signal plus a coincidence between the trigger scintillator and the Cherenkov counter. \textcopyright AAS. Reproduced with permission. }
    \label{fig:MPE}
\end{figure}

The most important result from this small detector  had nothing directly to do with $\gamma$-ray astrophysics.  Instead, during one balloon flight from Palestine, Texas, the instrument was pointed down to measure the upcoming atmospheric $\gamma$ radiation\cite{GSFC_6} at energies above 100 MeV.  This result showed a discrepancy with preliminary {\it OSO-3} atmospheric results.  The difference was large enough that the {\it OSO-3} team re-calibrated their flight spare unit, finding reason to adjust their measurements.  The final results from {\it OSO-3} reflect these changes\cite{OSO-3}.

Figure \ref{fig:MPE} is a schematic of another pair-production telescope with a similar design, also using a wire grid spark chamber\cite{MPE}. This instrument had thinner conversion plates, giving it sensitivity to $\gamma$ rays with energy down to 30 MeV. On two balloon flights from Palestine, Texas, this instrument gave indications of diffuse cosmic $\gamma$-ray emission in the 30-50 MeV energy range.  Measurement of any diffuse radiation in this energy range from a balloon is particularly difficult, because even at the highest balloon altitudes most of the detected photons are secondaries from charged particle cosmic rays interacting in the residual atmosphere.  The excess from extraterrestrial sources must be derived by subtracting the atmospheric component, based on the atmospheric flux as a function of altitude (the ``growth curve''). 

\subsection{Satellite Instruments}

The plethora of $\gamma$-ray telescopes based on pair production and carried on balloons ultimately demonstrated that the atmospheric background presented a nearly insurmountable challenge for all but the brightest sources in the sky.  The field would have to go to satellite instruments to make substantial progress.  The first generation of satellite high-energy  $\gamma$-ray telescopes was, not surprisingly, based on essentially the same technologies that had been developed and tested using balloon flights. 

A small spark chamber pair-production telescope was included in the {\it Orbiting Geophysical Observatory 5} payload, launched in 1968\cite{OGO-5}.  It used an acoustic readout of the spark data.  The high charged particle fluxes in orbit limited its capabilities, but it did see an excess of $\gamma$ rays above 40 MeV from the Cygnus region of the Galactic plane. 

The 1969 launch of {\it COSMOS-264} included a small $\gamma$-ray telescope with a spark chamber read out using film.  The limited supply of film restricted the results from this instrument. It did find an indication of E $>$ 100 MeV photons from the Active Galactic Nucleus 3C120\cite{COSMOS}. 

In early 1972, the {\it TD-1} satellite was launched, and among its instruments was the S-133 $\gamma$-ray telescope\cite{TD-1}. This detector used a vidicon camera system to read out the spark chamber data.  High background produced by cosmic-ray interactions with the surrounding instruments limited its usefulness, although it did detect one $\gamma$-ray burst\cite{TD-1_GRB}.

\begin{figure}[t]
    \centering
    \includegraphics[width=1.0\textwidth]{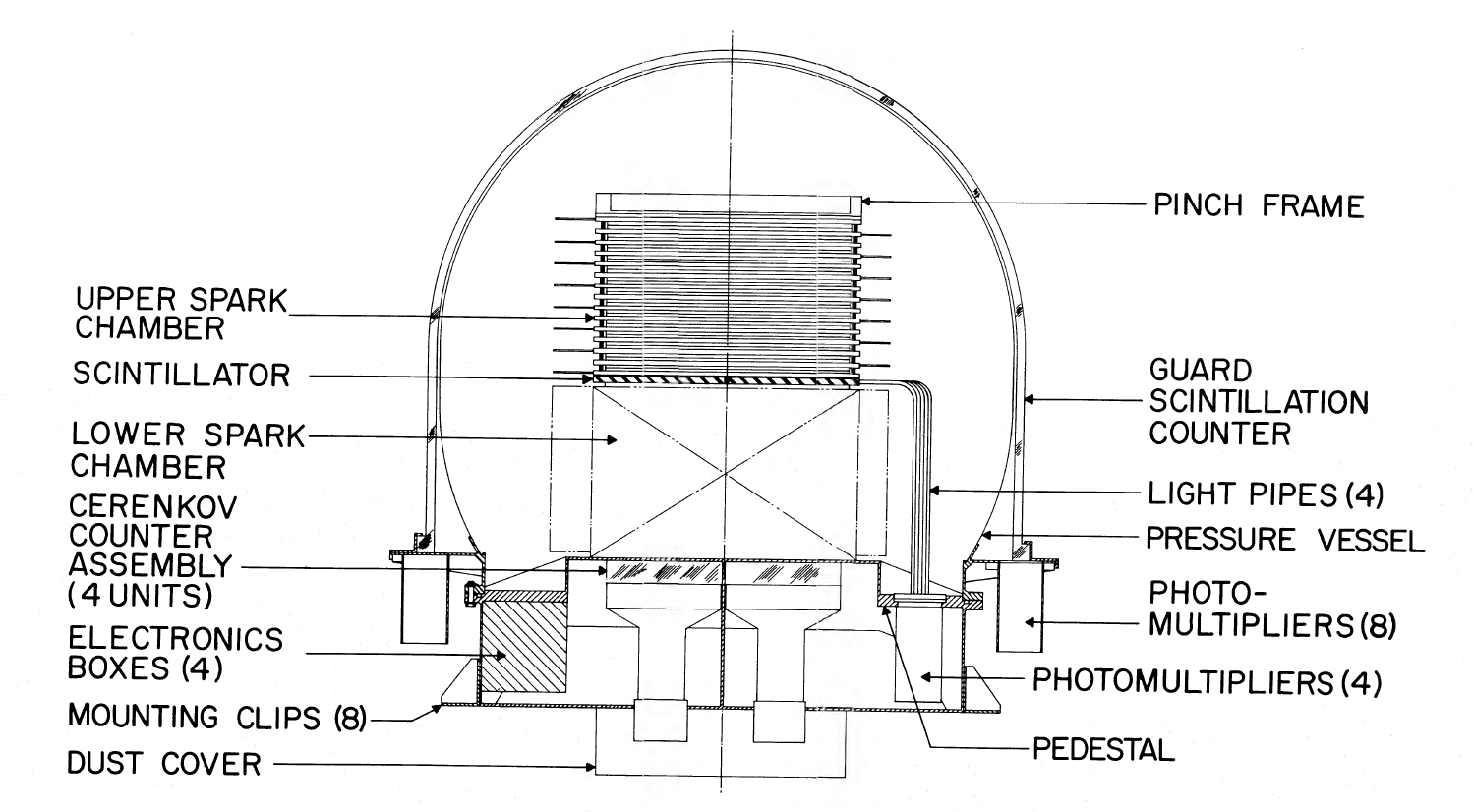}
    \caption{\small Schematic of the {\it Second Small Astronomy Satellite (SAS-2)} $\gamma$-ray telescope\cite{SAS-2}. \textcopyright AAS. Reproduced with permission. }
    \label{fig:SAS-2}
\end{figure}

The breakthrough satellite imaging $\gamma$-ray telescope was the {\it Second Small Astronomy Satellite (SAS-2)}, shown schematically in Fig. \ref{fig:SAS-2}\cite{SAS-2}.  The design was similar to some of the balloon instruments: the spark chamber used wire grids with magnetic core readout, and the triggering telescope used a mid-level scintillator and Cherenkov detector coincidence, with a plastic scintillator dome for anticoincidence. {\it SAS-2} had an effective area on axis of about 100 cm$^2$ and a field of view that extended to about 25$^\circ$ off axis. 

\begin{figure}[t]
    \centering
    \includegraphics[width=0.7\textwidth]{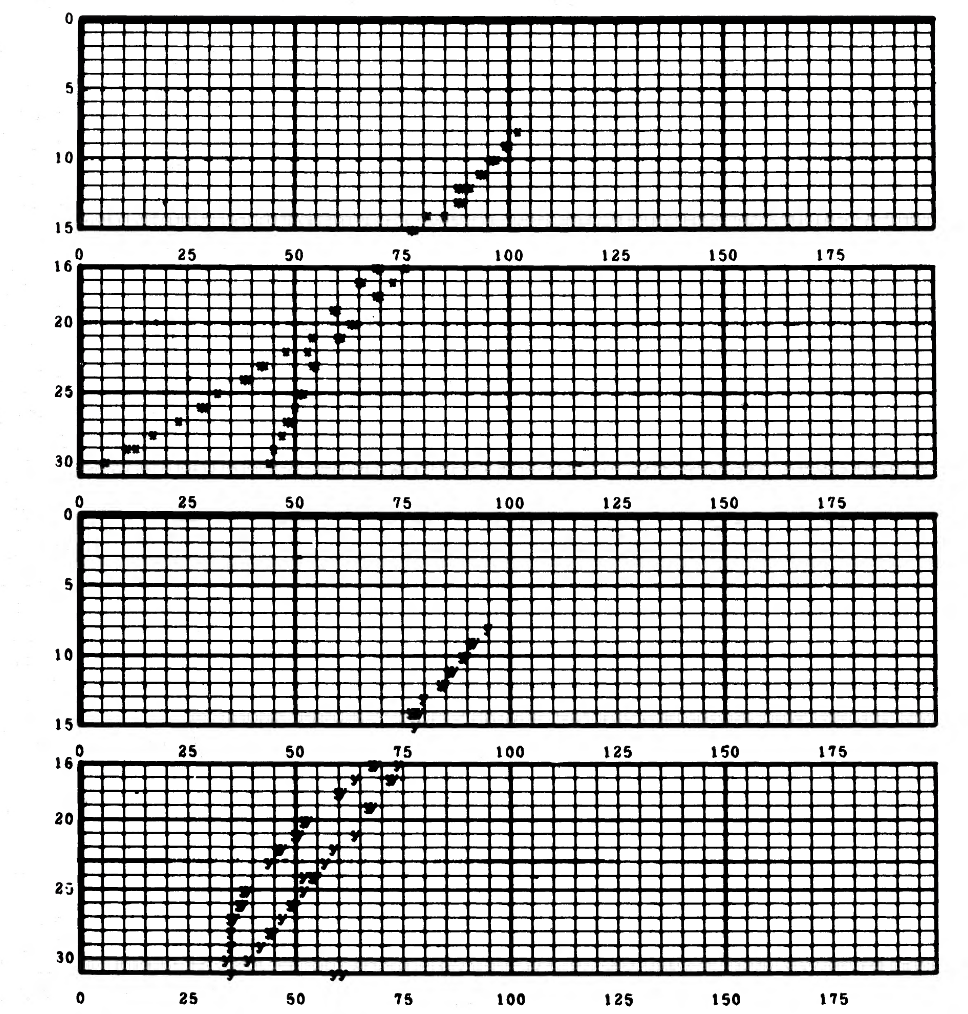}
    \caption{\small Orthogonal views of a {\it SAS-2} $ \gamma$-ray pair production event\cite{SAS-2}. \textcopyright AAS. Reproduced with permission.}
    \label{fig:SAS-2_event}
\end{figure}

A principal consideration for {\it SAS-2} was minimization of background, in order to produce the cleanest possible signal of cosmic $\gamma$ rays.  The discrimination against charged particles relied on the highly efficient, 32-layer spark chamber, where the appearance of a created pair produced a clear signature of a gamma ray (see Fig. \ref{fig:SAS-2_event}).  Outside the atmosphere, the $\gamma$-ray background originates from charged particles interacting in local material to produce $\gamma$-ray secondaries.  {\it SAS-2} used two approaches to minimize this effect: 

\begin{enumerate}
    \item {\it SAS-2} was launched in late 1972 on a Scout rocket from the Kenya launch facility.  This choice put the satellite into a low-Earth orbit very close to the equator, where the geomagnetic field  excludes the most cosmic-ray flux. 
    \item The inert material outside the anticoincidence scintillator was minimized and placed directly against the scintillator, so that any $\gamma$-ray-producing interactions had a high probability of being accompanied by a charged particle that could be detected by the anticoincidence system. 
\end{enumerate}

{\it SAS-2} operated in orbit for about six months, until an electronics failure ended the mission. It used one-week pointed observations to map much of the Galactic plane and a number of high-Galactic-latitude regions. Its effective energy range was 35 - 200 MeV, with the photon energies estimated from the multiple Coulomb scattering of the electron and positron. Key scientific results were:

\begin{itemize}
    \item The Galactic plane morphology and spectrum confirmed that $\gamma$ rays trace cosmic-ray interactions with the interstellar medium.
    \item The diffuse, isotropic, presumably extragalactic emission has a relatively steep energy spectrum\cite{SAS-2_diffuse}.
    \item The Crab and Vela pulsar $\gamma$-ray properties were measured\cite{SAS-2_Crab,SAS-2_Vela}.
    \item An unidentified source, which later came to be called Geminga, was discovered in the Galactic anticenter\cite{SAS-2_Geminga}. 
\end{itemize}

\begin{figure}[t]
    \centering
    \includegraphics[width=0.9\textwidth]{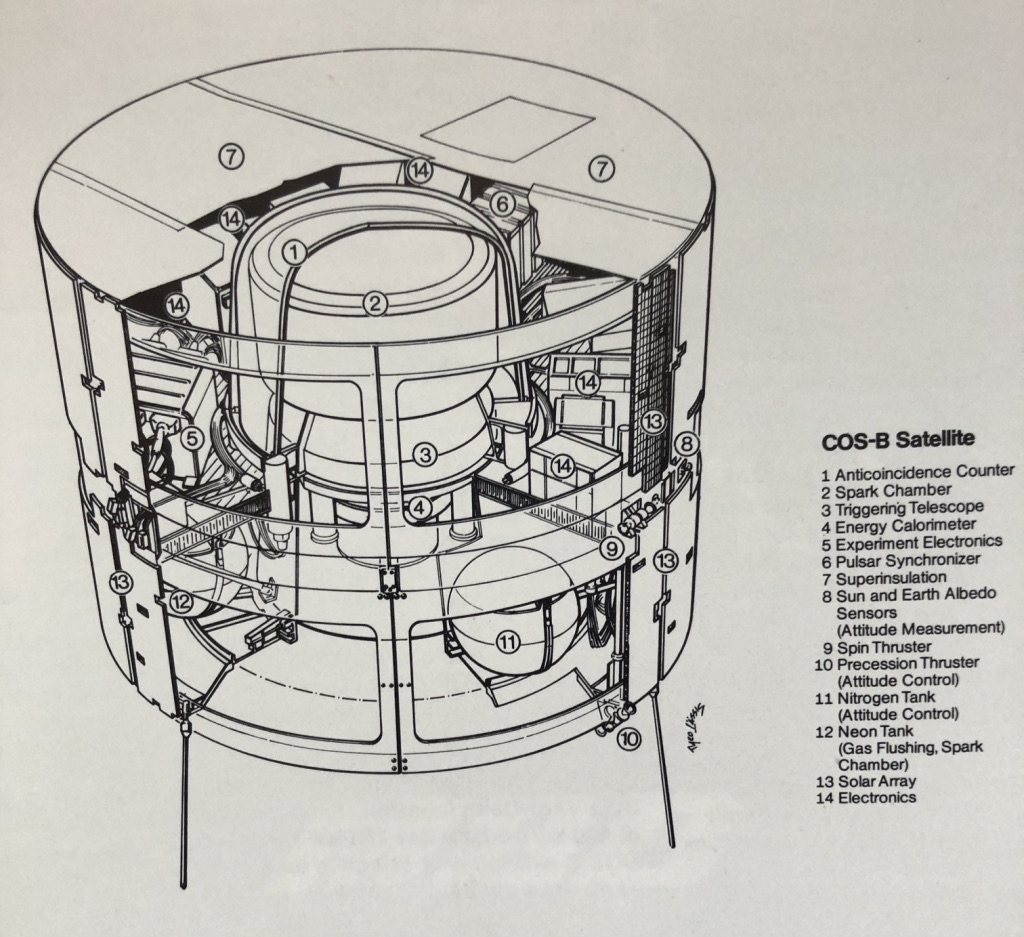}
    \caption{\small Schematic of the {\it COS-B} $\gamma$-ray telescope. (Credit: Messerschmitt-B\"olkow-Blohm GmbH brochure, URB 5-7-75-25 D, 1975) }
    \label{fig:COS-B}
\end{figure}

The {\it COS-B} satellite\cite{COS-B}, launched in 1975, provided another major advance for astrophysics in the pair-production energy regime. A schematic is shown in Fig. \ref{fig:COS-B}. Although it was nearly the same size as {\it SAS-2} and also had a wire grid spark chamber as a primary detector, {\it COS-B} had a number of differences:

\begin{itemize}
    \item The triggering telescope had three coincidence elements: two scintillators and a Cherenkov detector.
    \item {\it COS-B} included a CsI calorimeter to measure the energies of the electron-positron pair, giving it a much broader effective energy range (30 MeV - 10 GeV) than {\it SAS-2}.
    \item Because spark chamber gas performance deteriorates with use, {\it COS-B} carried a supply of replacement gas, allowing it to operate successfully for more than 6 years. 
    \item {\it COS-B} had structural material above the anticoincidence scintillator, and the orbit was an elongated one that exposed the instrument to the full flux of charged-particle cosmic rays. As a result, {\it COS-B} suffered from some locally generated $\gamma$-ray background. 
\end{itemize}

{\it COS-B} operations consisted of a series of pointed observations, concentrated mostly, but not entirely, on the Galactic plane, where the bright diffuse Galactic emission mitigated against the instrumental background. The {\it COS-B} longevity and its broad energy range provided a much more extensive view of the $\gamma$-ray sky than was obtained by {\it SAS-2}. Some scientific highlights were:

\begin{itemize}
    \item {\it COS-B} produced the first real catalog of high-energy $\gamma$-ray sources\cite{2CG}. Most of the 25 sources in the 2CG catalog did not have obvious counterparts at other wavelengths. 
    \item Quasar 3C273 was the first high-confidence detection of an extragalactic high-energy  $\gamma$-ray source\cite{COS-B_3C273}.
    \item The Orion cloud complex was mapped in $\gamma$ rays\cite{COS-B_Orion}.
    \item Detailed analysis was possible for the Galactic diffuse emission and sources seen by {\it SAS-2}.
\end{itemize}

\section{Second Generation Imaging Detectors}
\label{Second_Generation}

During the time {\it SAS-2} and {\it COS-B} were operating, other pair-production $\gamma$-ray telescopes were being planned, built, and tested.  These included larger instruments and different experimental techniques from the earlier generation.  Some were flown on balloons, often as prototypes of potential future satellite missions.  The most successful satellite instrument from this period was the Energetic Gamma-Ray Experiment Telescope (EGRET) on the {\it Compton Gamma Ray Observatory}.

\subsection{Advanced Balloon Instruments}

A larger version of a wire-grid magnetic core spark chamber instrument flown on balloons provided confirmation of the {\it OSO-3} Galactic Center region emission\cite{Half-meter}. This same instrument was later reconfigured with thinner pair production conversion plates to reduce its energy threshold to about 15 MeV, and balloon flight data measured the Galactic emission at lower energies than seen previously\cite{Half-meter-thin}.

\begin{figure}[t]
    \centering
    \includegraphics[width=0.9\textwidth]{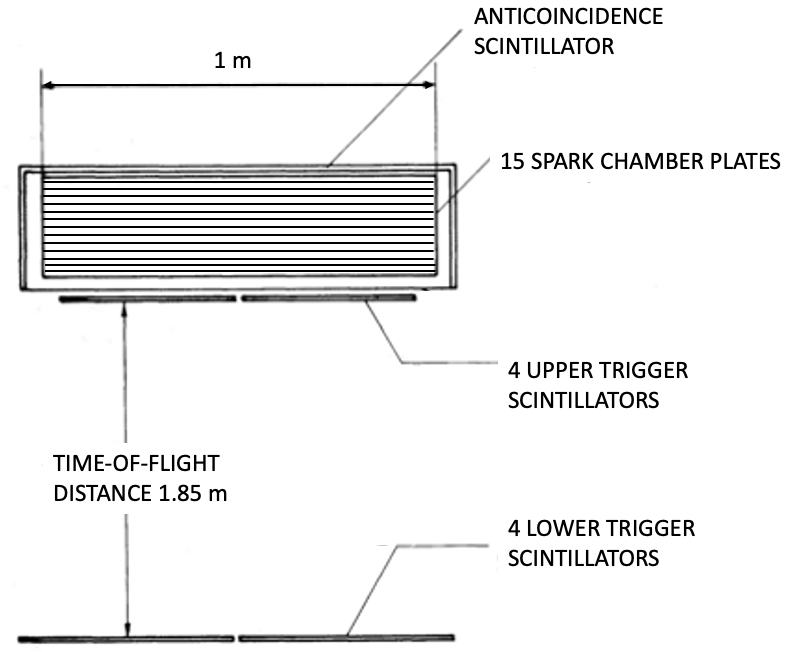}
    \caption{\small Schematic of the Agathe $\gamma$-ray telescope, adapted from \cite{Agathe}. }
    \label{fig:Agathe}
\end{figure}

A similar approach of using thin conversion plates to reduce the energy threshold was applied to another of the balloon-borne spark chamber telescopes, yielding a hint of pulsed 10$-$30 MeV emission from the Vela pulsar\cite{Albats_Vela}. An even larger (1 square meter) balloon payload using multi-wire proportional chambers instead of spark chambers introduced a new technology while maintaining the same basic pair production configuration\cite{Frye_MWPC}.  Balloon flights of this telescope yielded a pulsed signal from the Crab pulsar in the 10$-$30 MeV energy range. 

Another technical innovation was introduced with the Agathe $\gamma$-ray telescope\cite{Agathe}. Figure \ref{fig:Agathe} shows a schematic of this instrument.  In addition to using large area (1 square meter) optical spark chambers with very thin plates, Agathe used a time-of-flight triggering system to discriminate against upward-moving particles, instead of the Cherenkov detectors used on most previous instruments. Balloon flights from Brazil produced measurements of the atmospheric and diffuse $\gamma$-ray flux in the energy range 4$-$25 MeV. 

A wire grid spark chamber instrument with a time-of-flight triggering system was flown from Palestine, Texas, as a demonstration of technology for a satellite mission\cite{HEBE}. Data analysis included an automated pattern recognition system used to screen out unwanted triggers.  Performance matched expectations, confirming the usefulness of the design. 

Another satellite prototype using wide-gap spark chambers and time-of-flight triggering was also designed for balloon testing\cite{Voronov}.  A unique feature of this instrument was the use of position-sensitive scintillator strips to trigger the spark chambers only when events arrived close to the direction of a pre-defined target.

\begin{figure}[t]
    \centering
    \includegraphics[width=1.0\textwidth]{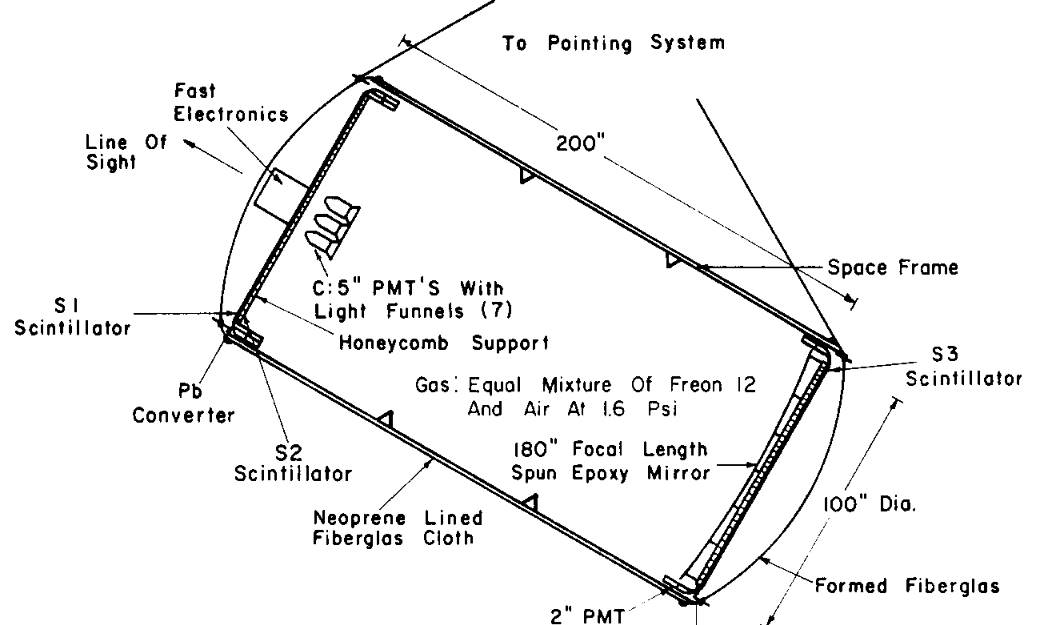}
    \caption{\small Schematic of the large gas Cherenkov $\gamma$-ray telescope \cite{Gas_Cher}. Reproduced with permission.}
    \label{fig:Gas_Cherenkov}
\end{figure}

A completely different approach to a pair production $\gamma$-ray telescope was based on a large gas Cherenkov instrument\cite{Gas_Cher}. A schematic of this balloon instrument is shown in Fig. \ref{fig:Gas_Cherenkov}.  A high-energy  $\gamma$ ray entering along the line of sight (upper left in the figure) produces no signal in the anticoincidence scintillator, then undergoes pair production in the thin lead converter.  The electron and positron trigger the S2 scintillator, then produce Cherenkov light while moving through the Freon gas.  The light is reflected from the spherical mirror segment at the rear and is detected by the cluster of phototubes C. The time of flight between the S2 signal and the C signal is 32 nsec. The field of view was about 7$^{\circ}$ in diameter,  and the arrival direction of the photons could be measured to about 0.5$^{\circ}$. The energy threshold was about 100 MeV, although its maximum efficiency came at energies above 700 MeV. In flights from Palestine, Texas, in 1971 and 1973, this instrument clearly detected pulsed emission from the Crab Pulsar, with indications of a change in the pulsed flux between the two flights\cite{Gas_Cher_1971,Gas_Cher_1973}.

\subsection{Second-generation Imaging Satellite Instruments}

By the late 1970's, the particle physics community had largely shifted from spark chambers to multiwire proportional chambers or drift chambers.  Nevertheless, two satellite pair production telescopes that were approved in this time frame in the USA (EGRET) and in the Soviet Union and France (Gamma-1) used spark chambers as the main detectors. In addition to space-flight experience and high reliability, spark chambers could be operated using much less power than these other technologies, and minimizing power is always a consideration in space. 

\begin{figure}[t]
    \centering
    \includegraphics[width=0.7\textwidth]{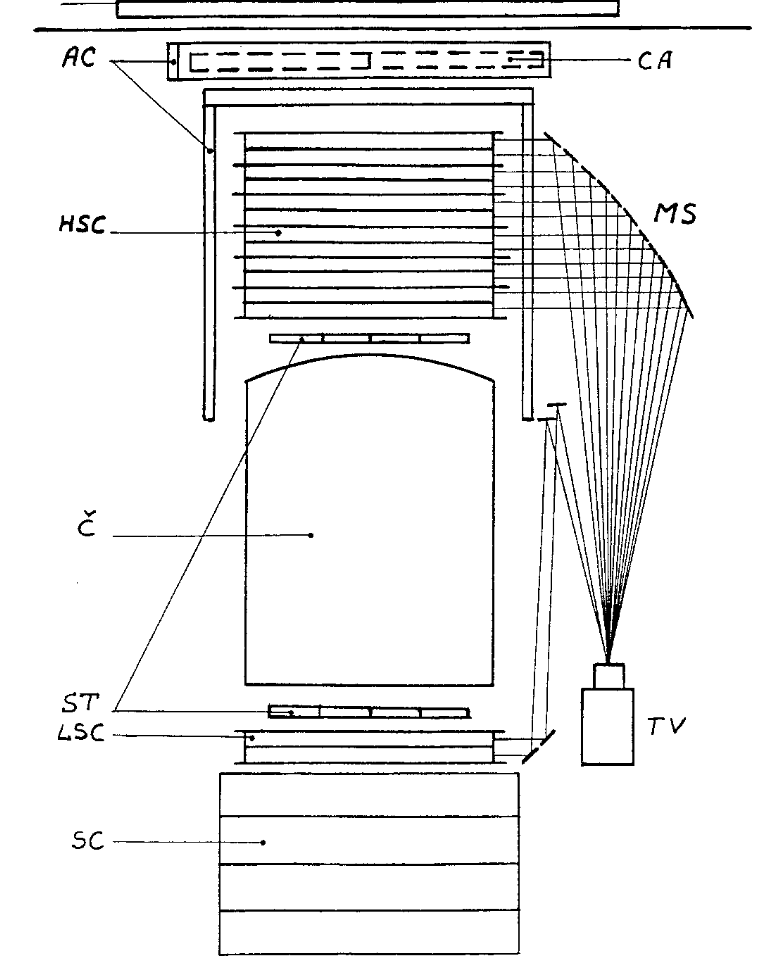}
    \caption{\small Schematic of the Gamma-1 $\gamma$-ray telescope on the {\it Gamma} satellite\cite{Gamma-1}. AC: anticoincidence scintillator; CA: coded mask (planned but not flown); HSC and LSC: spark chambers; MS: mirrors; ST: trigger scintillators; C: Cherenkov detector; SC: scintillator calorimeter.  Reproduced with permission.}
    \label{fig:Gamma-1}
\end{figure}

Gamma-1\cite{Gamma-1} was conceived  based on an extended history of space-borne and balloon-borne small gamma-ray telescopes.
The Gamma-1 main detector was a 12-layer stack of wide-gap optical spark chambers with geometrical area 50~cm $\times$ 50~cm (Fig. \ref{fig:Gamma-1}). 
The spark chambers were viewed by a mirror system and the tracks were recorded  by a vidicon system. The spark chambers were triggered by the coincidence of the signals of two time-of-flight plastic scintillators and the gas Cherenkov detector placed in between them. This trigger arrangement provided high rejection of upward-moving particles, and also additional rejection of cosmic ray protons with energy below the Cherenkov detection threshold of about 12 GeV. The stack of spark chambers was surrounded by efficient anticoincidence plastic scintillators to detect and veto charged cosmic rays. The 7.4 radiation length thick scintillation calorimeter was placed at the bottom of the telescope to measure the energy of detected photon.

The telescope operated in the energy range from 50 MeV to 5 GeV, with angular resolution $\sim$ 2$^{\circ}$ at   100 MeV, improving to $\sim$ 1.2$^{\circ}$ at 300 MeV. Its energy resolution was 70\% at 100 MeV, improving to 35\% at 550 MeV. Originally  Gamma-1 was planned to have a coded-aperture mask which would dramatically improve its angular resolution, but by launch time it was abandoned due to significant complication of the telescope design. 

Gamma-1 on the {\it Gamma} satellite was launched from Baikonur on July 11, 1990 and suffered from a major problem - for a still unknown reason power was not delivered to the spark chambers, so the telescope angular resolution was provided only by the Cherenkov detector at a level of $\sim$ 12$^{\circ}$. Two years after the launch it was decided to terminate this ill-fated mission. During the orbital operation Gamma-1 performed observations of Vela pulsar, the Galactic Center region, the Cygnus X-1 and X-3 binaries, and especially Her X-1. Interesting information was obtained about the high-energy emission of the Sun during peak solar activity, including solar flares\cite{Gamma-1_solar}.

\begin{figure}[t]
    \centering
    \includegraphics[width=1.0\textwidth]{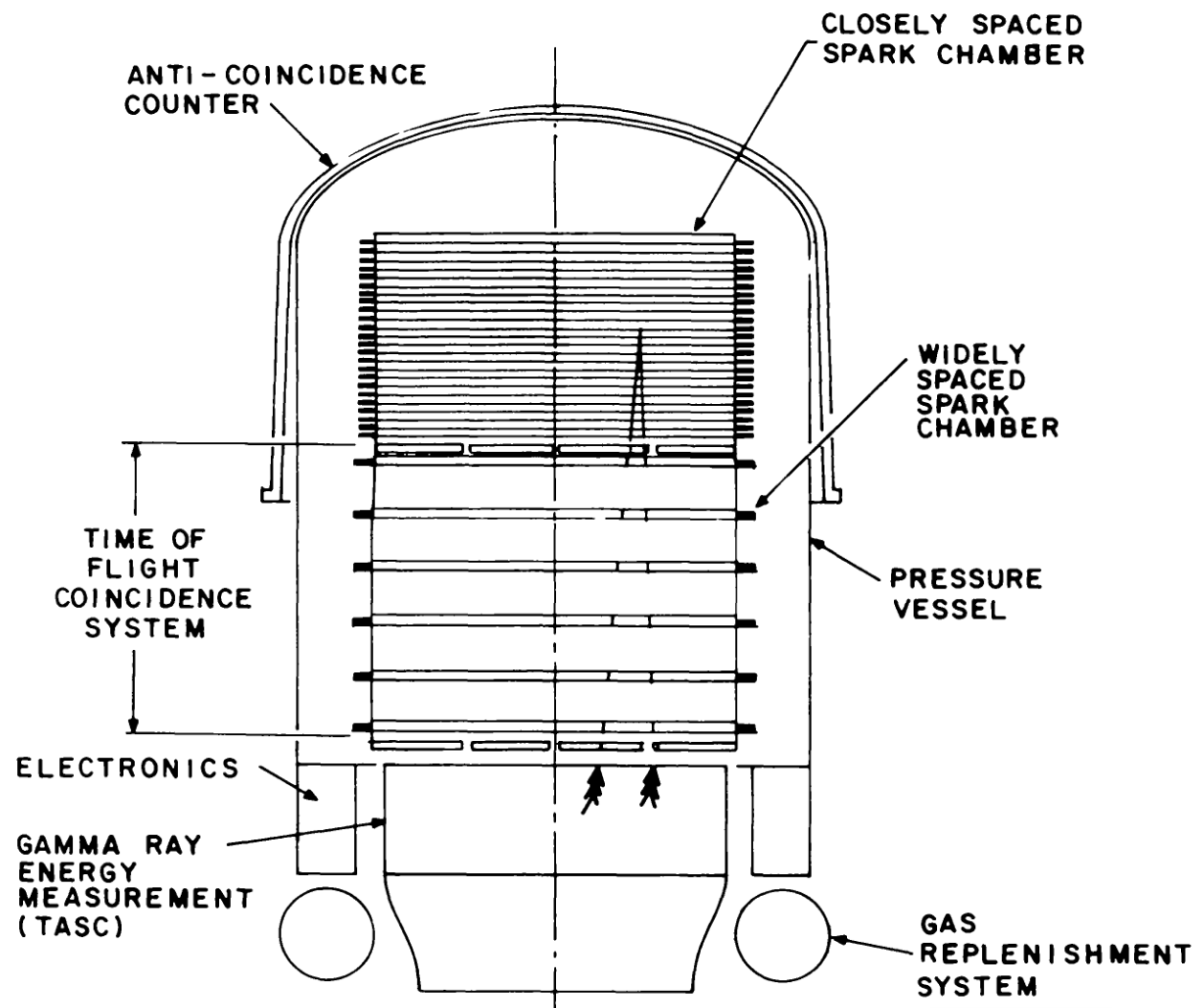}
    \caption{\small Schematic of the Energetic Gamma Ray Experiment Telescope (EGRET) on the {\it Compton Gamma Ray Observatory}\cite{EGRET}. Reproduced with permission.}
    \label{fig:EGRET}
\end{figure}

NASA's choice for a pair production telescope for the {\it Gamma Ray Observatory}, later renamed the {\it Compton Gamma Ray Observatory (CGRO)}, was the Energetic Gamma Ray Experiment Telescope (EGRET).  As part of the Great Observatories series, {\it CGRO} carried a suite of $\gamma$-ray telescopes, so that resources like mass and power on the satellite had to be shared. EGRET, shown schematically in Fig. \ref{fig:EGRET} was in many ways a conventional instrument, incorporating the best features of previous satellite and balloon pair production telescopes\cite{EGRET}.  Key elements of the design included:

\begin{itemize}
    \item A wire-grid, magnetic-core spark chamber interleaved with thin tantalum foils served as a converter and tracker for the pair production events.
    \item A time-of-flight coincidence system, operated in anticoincidence with the overlying scintillator dome formed the trigger. 
    \item An 8-radiation-length Sodium Iodide crystal system provided energy measurements.  A minimum signal in this calorimeter was also incorporated into the trigger configuration. 
    \item A gas replenishment system enabled refills of the spark chamber gas as it aged. 
\end{itemize}
Table 1 shows some of the operational properties of EGRET, based on extensive accelerator calibrations\cite{EGRET_Calibration}. 

The satellite carrying EGRET was placed into a low-Earth orbit by the Space Shuttle Atlantis on 5 April 1991. Regular science operations started on 15 May 1991.  {\it CGRO} was operated in a series of pointings with the goal of surveying the $\gamma$-ray sky.  After about 18 months, EGRET completed the first all-sky map of the sky at energies above 100 MeV. Although originally planned as a 2-year mission, {\it CGRO} operated in space for over 9 years.  Following the failure of part of the attitude control system, the satellite was de-orbited. EGRET continued to operate through the end of the mission. 

\vspace{5mm} 
{\begin{tabular}{@{}ll}
Table 1\\
Some Characteristics of EGRET\\\\
Property&Value\\
\hline
Energy Range&20 MeV $-$ 30 GeV\\
Peak Effective Area&1500 cm$^2$ at 500 MeV \\
Energy Resolution&15\% FWHM\\
Effective field of view&0.5 steradians\\
Timing accuracy&$<$ 100 $\mu$sec absolute\\
Point spread function&$\sim$6 deg. 68\% containment at 100 MeV\\
\hline
\end{tabular}}
\vspace{5mm} 

EGRET's scientific results cover a broad range of high-energy astrophysics in the pair production energy regime\cite{EGRET_Summary}.  A few of the highlights derived from mapping and monitoring the full $\gamma$-ray sky are:

\begin{itemize}
    \item The Third EGRET Catalog (3EG) contained 271 sources, over half of which were not clearly associated with known astrophysical objects\cite{3EG}. Some of these sources were later eliminated by the discovery of ``dark gas''\cite{Grenier}. 
    \item At least five $\gamma$-ray pulsars were found, including the discovery that Geminga is a radio-quiet pulsar\cite{Geminga}.
    \item The detection of the Large Magellanic Cloud as a spatially extended $\gamma$-ray source was the first normal galaxy seen in this energy range\cite{EGRET_LMC}. 
    \item The largest identified source class seen by EGRET was blazar-type active galactic nuclei, with many of these showing strong variability in the $\gamma$-ray energy range\cite{EGRET_AGN}, enabling multiwavelength correlation studies.
    \item For the first time, high-energy  $\gamma$-ray activity was observed from a $\gamma$-ray burst\cite{EGRET_GRB}. 
\end{itemize}

In addition to results like these, the EGRET observations yielded hints of other $\gamma$-ray sources, including supernova remnants, pulsar wind nebulae, OB associations, X-ray binaries, and radio galaxy Cen A.  

\section{Third Generation - Solid State Imaging Detectors}
\label{Solid_State}

By the time the Gamma-1 and EGRET pair production telescopes had been launched, the particle physics community had made significant progress in particle tracking detectors, which form the central element of any high-energy $\gamma$-ray telescope.  In particular, the advent of solid state detectors with sufficiently low power requirements to be practical for space applications enabled instruments that circumvented the limitations of the gas detectors used previously. At the time of this writing, such telescopes represent the workhorses for pair production $\gamma$-ray telescopes.

\begin{figure}[t]
    \centering
    \includegraphics[width=1.0\textwidth]{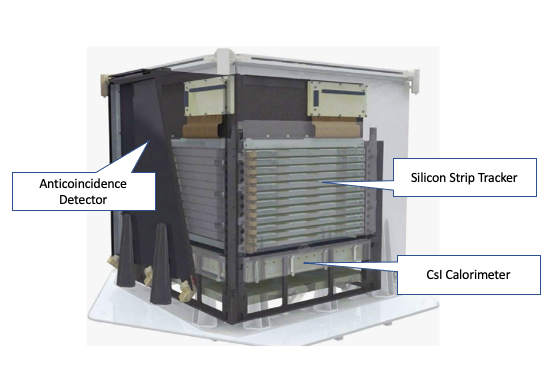}
    \caption{\small {\it Astro-rivelatore Gamma a
Immagini LEggero (AGILE)}\cite{AGILE}. The picture shows a cutaway of the outer anticoincidence scintillator along with the silicon strip tracker and the thin CsI calorimeter that make up the Gamma-Ray Imaging Detector (GRID), adapted from \cite{AGILE}. Reproduced with permission. }
    \label{fig:AGILE}
\end{figure}

Although its emphasis was on $\gamma$ rays interacting by Compton scattering, the Medium Energy Gamma-ray Astronomy (MEGA) concept envisioned using silicon-strip detectors as a tracker that could also be used in a pair production mode \cite{MEGA}. By using double-sided silicon-strip detectors and eliminating the converter foils of previous telescopes, MEGA could lower the energy threshold for pair-production $\gamma$-ray detection. A prototype of MEGA was built and calibrated, but not flown.  The MEGA concept, however, has been adapted for some proposed Compton/pair instruments (see section \ref{Future}), and the MEGAlib software developed for the instrument \cite{MEGAlib} has become a standard tool for simulating medium-energy $\gamma$-ray telescopes.

The first of the new generation of pair production telescopes to fly was the small satellite {\it Astro-rivelatore Gamma a Immagini LEggero (AGILE)}\cite{AGILE}, launched into a low-Earth, low-inclination orbit in 2007 from the Satish Dhawan Space Centre in India. A cutaway view of the active parts of the instrument is shown in Fig. \ref{fig:AGILE}.  The anticoincidence scintillator to reject charged particles and the CsI calorimeter for energy measurement are similar to past missions.  The important new development is the use of a silicon strip tracker, interleaved with tungsten foils for pair conversion, as the heart of the Gamma-Ray Imaging Detector (GRID) part of {\it AGILE}. The AGILE silicon strips have a readout pitch of 242 $\mu$m, and its analog readout enables far more accurate measurements of the electron-positron tracks than was possible with spark chambers. Silicon strip detectors, read out by Application Specific Integrated Circuits (ASICs), can not only provide accurate locations of particle tracks, they can also self-trigger, eliminating the need for trigger or time-of-flight scintillators. Some operating characteristics of  the {\it AGILE} GRID are shown in Table 2. 

\vspace{5mm} 
{\begin{tabular}{@{}ll}
Table 2\\
Some characteristics of the {\it AGILE} GRID\\\\
Property&Value\\
\hline
Energy Range&30 MeV $-$ 50 GeV\\
Peak Effective Area&$\sim$500 cm$^2$ above 1000 MeV \\
Energy Resolution&$\sim$50\% FWHM\\
Effective field of view&2.5 steradians\\
Timing accuracy&2 $\mu$sec absolute\\
Point spread function&$\sim$4 deg. 68\% containment at 100 MeV\\
\hline
\end{tabular}}
\vspace{5mm} 

{\it AGILE} was operated originally in pointing mode, taking advantage of its large field of view to observe large portions of the sky in each viewing.  In late 2009 a portion of the satellite attitude control system failed, and the operation shifted to a spinning mode with a 7-minute rotation period, during which the GRID obtains exposure to nearly 80\% of the sky. As of this writing, {\it AGILE} remains active. 

Scientific results from {\it AGILE} GRID span a wide range of topics, particularly focused on transient and other time-variable sources.  Some highlights of these results include:

\begin{itemize}
    \item The Second {\it AGILE} Catalog (2AGL) contains 175 sources, about 30\% of which are not clearly associated with known astrophysical objects\cite{2AGL}. 
    \item Strong flares were discovered from the Crab Nebula at energies above 100 MeV\cite{AGILE_Crab}.
    \item Evidence was found that $\gamma$-ray activity from Cygnus X-3 is related to changes in the hard X-ray flux\cite{AGILE_Cyg}. 
    \item The $\gamma$-ray energy spectrum of supernova remnant W44 shows curvature that indicates hadronic cosmic-ray particle acceleration\cite{AGILE_W44}.
\end{itemize}

\begin{figure}[t]
    \centering
    \includegraphics[width=1.0\textwidth]{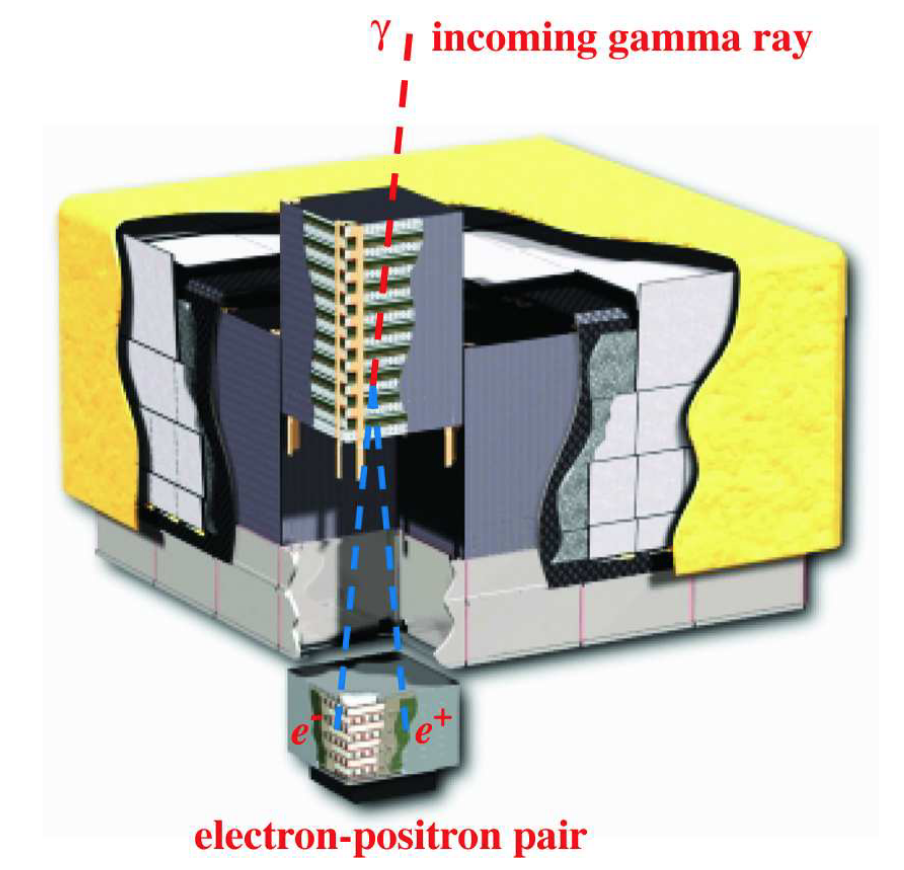}
    \caption{\small {\it Fermi Gamma-ray Space Telescope} Large Area Telescope (LAT)\cite{Atwood_LAT}. \textcopyright AAS. Reproduced with permission. }
    \label{fig:LAT}
\end{figure}

From the time of its launch into a low-Earth orbit and activation in the summer of 2008, the Large Area Telescope (LAT) on the {\it Fermi Gamma-ray Space Telescope} (formerly GLAST) has been the most sensitive $\gamma$-ray telescope operating in the pair production regime\cite{Atwood_LAT}. The LAT uses similar technology to that of the {\it AGILE} GRID, but on a much larger scale (Fig \ref{fig:LAT}).  The primary detector subsystems are:

\begin{itemize}
\item The tracker is a stack of 18 x-y pairs of single-sided silicon strip detectors (228 $\mu$m pitch, without analog readout), with the first 16 interleaved with tungsten converter foils.  The trackers are modular, with 16 towers in a 4 $\times$ 4 array.  In the primary trigger mode, the tracker self-triggers on a pattern of three consecutive x-y pairs producing a signal. 
\item  The energy measurement is carried out by the calorimeter consisting of 1536 CsI logs,  in 8 layers with directions alternating x and y, read out by custom photodiodes. The calorimeter modules are arranged in the same 4 $\times$ 4 array of towers  as the tracker.  
\item   The LAT anticoincidence system consists of 89  plastic scintillator tiles, overlapped, with scintillating fibers covering seams.   The signals of charged particles are measured by photomultiplier tubes. 
\end{itemize}

LAT triggers, including the primary trigger (the tracker self-trigger + no signal in the anticoincidence detector + a minimum signal in the calorimeter) and a number of auxiliary ones, generate information from each of the subsystems.  Onboard analysis screens the data to select those events most likely to provide useful scientific data.  These events are transmitted to the ground. 

Table 3 show some of the important performance properties of the LAT. The LAT has experienced no significant hardware failures in its more than 13 years of operation. Thanks to improvements in data processing, the LAT scientific performance is now better than it was at the time of launch \cite{LAT_10}.

\vspace{5mm} 
{\begin{tabular}{@{}ll}
Table 3\\
Some characteristics of the {\it Fermi} Large Area Telescope\\\\
Property&Value\\
\hline
Energy Range&20 MeV $-$ $>$300 GeV\\
Peak Effective Area&8000 cm$^2$ above 10 GeV \\
Energy Resolution&$<$15\% FWHM\\
Effective field of view&2.4 steradians\\
Timing accuracy&$<$ 1 $\mu$sec absolute\\
Point spread function&$\sim$5 deg. 68\% containment at 100 MeV\\
\hline
\end{tabular}}
\vspace{5mm} 

{\it Fermi}  has been operated  largely in a scanning mode since its launch, taking advantage of the instrument's huge field of view to monitor the sky continuously \cite{Fermi_mission}. By rocking the pointing direction north and south of the orbital plane on alternating 96-minute orbits, the LAT can view the entire sky every $\sim$3 hours. The {\it Fermi} spacecraft had a failure in 2018 March:  one of the solar panel drive motors failed, leaving that panel unable to track the Sun.  The solar array can provide full power to the instruments and spacecraft, but this limitation has required a partial change of observing strategy.  Instead of viewing the full sky every three hours most of the time, the LAT sometimes needs a few weeks to produce an all-sky image.

Scientific results from the {\it Fermi} LAT span a broad range of subjects. Here are a few of the highlights: 

\begin{itemize}
    \item The Fourth {\it Fermi} LAT catalog \cite[4FGL][]{4FGL} contains 5064 sources, about one-quarter of which are not clearly associated with known astrophysical objects. 
    \item  The {\it Fermi} Bubbles, discovered in 2010 using the public {\it Fermi} data \cite{bubbles}, are huge $\gamma$-ray-emitting features extending about 12 kpc above and below the direction to the Galactic Center (assuming that distance).
    \item {\it Fermi}-LAT detections of classical and recurrent novae demonstrate that shocks from the explosions in these binary systems can accelerate particles to high enough energies to produce $\gamma$ rays \cite{novae_classical}. 
    \item Some models of Dark Matter (DM) based on Weakly Interacting Massive Particles (WIMPs), predict detectable  $\gamma$-ray emission resulting from WIMP decay or annihilation. Absence of $\gamma$-ray emission from dwarf spheroidal galaxies, with a strong DM component and none of the usual types of $\gamma$-ray  sources, put strong constraints on WIMP models \cite{dwarfs}. Excess $\gamma$-ray emission from the Galactic Center region, however, might  indicate DM $\gamma$ radiation \cite[e.g.,][]{DM_GC}. 
    \item LAT has discovered more than 250 $\gamma$-ray pulsars, including  radio-quiet pulsars, millisecond pulsars, pulsars whose $\gamma$-ray flux changes significantly, pulsars that transition between rotation-powered emission and Low-Mass X-ray Binary states, and pulsars in binary systems with eclipses that allow accurate measurements of neutron star masses\cite{pulsar_summary}.
    \item Active Galactic Nuclei (AGN), especially blazars with  powerful jets pointed close to the line of sight, are the dominant source class seen by the LAT and are excellent probes of many astrophysical phenomena. The {\it Fermi}-LAT $\gamma$ rays can interact with optical or infrared photons in an annihilation process; therefore absorption signatures in blazar $\gamma$-ray spectra can be used to measure the extragalactic background light and the star-formation history of the universe \cite{star-formation}. A flaring LAT blazar also provided the first solid evidence of a source of high-energy neutrinos\cite{TXS_neutrino}.
    \item At least 40 LAT sources are associated with supernova remnants, and some of these have energy spectra that show the low-energy cutoff characteristic of neutral pion decay, a clear signature of hadronic interactions \cite{SNR}.
\end{itemize}

Two other operating space missions have pair production detection capability. Although the {\it Alpha Magnetic Spectrometer 2 (AMS-02)} and the {\it DArk Matter Particle Explorer (DAMPE)} are primarily operated as particle detectors, they each have particle tracking capability that enables $\gamma$-ray detection. {\it AMS-02}, located on the International Space Station, has two ways of detecting pair production events\cite{AMS}: (1) The silicon strip tracker, located within the magnetic field, can detect 100 MeV - 10 GeV $\gamma$ rays converting in the top part of the instrument.  The trigger uses the time-of-flight system. (2) Higher-energy photons (1 - 100 GeV) can be detected within the lead-scintillating fiber sampling calorimeter near the bottom of the instrument.  The sensitivity of {\it AMS-02} is much smaller than that of {\it Fermi} LAT, but results have been obtained for both diffuse emission and individual sources.  {\it DAMPE}, in a low-Earth orbit, detects pair production events from $\gamma$ rays (1 - 100 GeV) that convert in the upper part of the instrument, produce tracks in the silicon strip tracker, and have energy measured in the Bismuth Germanate calorimeter\cite{DAMPE}. Timing results for some $\gamma$-ray pulsars and flares from AGN have been reported from {\it DAMPE}.

\section{Continuing Developments and the Future}
\label{Future}

With four pair production telescopes currently operating in space,  the field of high-energy $\gamma$-ray astrophysics has shown remarkable progress in recent years, led by the {\it Fermi} LAT's thousands of new $\gamma$-ray sources.  Future progress in this field will strive to address some aspects of the current missions that offer room for improvement, including sensitivity (important for time variable sources), energy range (especially at energies below 100 MeV), angular resolution (critical for source detection and mapping), and polarization (a valuable tool for understanding physical processes). A variety of technical approaches have been suggested. 

The Gamma-Ray Astro-Imager with Nuclear Emulsion (GRAINE) instrument is a balloon-borne telescope using emulsions, including a stack shifter that allows time stamps with 1 second accuracy\cite{GRAINE}.  Emulsions allow high angular resolution ($\sim$ 1$^{\circ}$ at 100 MeV) and polarization measurements, and the size can be scaled up easily.  Results have measured the atmospheric $\gamma$ rays and shown the ability to detect sources. 

Extending the useful energy range for pair production detection downward toward 10 MeV requires less scattering material in the particle tracking subsystem, because the limitation has been the relatively poor angular resolution of instruments like {\it Fermi} LAT at energies below 100 MeV. One approach toward this goal is to develop a detector similar to the LAT but with only double-sided silicon strips as pair production converter material.  The proposed  e-ASTROGAM\cite{ASTROGAM}, and All-sky Medium Energy Gamma-ray Observatory (AMEGO)\cite{AMEGO} telescopes would use this technique. Similar to the MEGA concept (see section \ref{Solid_State}), these instruments would use both pair production and Compton scattering for detection. The PAir-productioN Gamma-ray Unit (PANGU)\cite{PANGU} concept is similar but with single-sided silicon strip detectors. 

Alternatively, lower-energy good angular resolution and polarization can be achieved by using a gaseous time projection chamber as a pair production tracker instead of silicon.  An example is the Advanced Energetic Pair Telescope (AdEPT)\cite{ADEPT}.  The pair production takes place in a gas volume, and an applied voltage causes the electron and positron to drift toward a 2-dimensional micro-well detector array.  The drift time provides the third dimension. Accelerator tests of a prototype of a similar detector, Hermetic ARgon POlarimeter (HARPO)\cite{HARPO} have demonstrated good angular resolution and capability to measure polarization at energies down to 12 MeV. 

Yet another approach to tracking pair production events could be the use of scintillating fibers.  This tracking technique can in principle be scaled up in size more easily than other methods, potentially enabling higher $\gamma$-ray sensitivity.  Such an instrument was proposed but not selected for the GLAST mission\cite{FiberGLAST}. Possible future missions based on scintillating fibers include the Advanced Particle-astrophysics Telescope (APT)\cite{APT}, High Energy cosmic-Radiation Detection (HERD) facility\cite{HERD}, and Gamma-400\cite{G400}. 

At present, no pair-production $\gamma$-ray satellite instrument is under construction.  The future of this field therefore remains an open question. 

\section{Summary}

The use of pair production ($\gamma \rightarrow e^- + e^+$) is required for $\gamma$-ray astrophysical measurements at photon energies above a few tens of MeV.  Balloon and satellite telescopes using this process have evolved over the past 60 years, drawing heavily on particle detection instrumentation used in accelerator physics.  The current state-of-the-art $\gamma$-ray telescopes, {\it AGILE} and {\it Fermi} LAT, continue to explore a dynamic, high-energy universe, with a broad range of astrophysical discoveries having emerged. Multiwavelength and multimessenger studies continue to be a focus of observations in the pair production energy range. 

\section{Cross References}

\begin{itemize}
    \item Telescope concepts in gamma-ray astronomy
    \item Orbits and background of gamma-ray space instruments
    \item Silicon detectors for gamma-ray astronomy
    \item Scintillators for gamma-ray astronomy
    \item Detector and mission design simulations
    \item The Large Area Telescope on the Fermi mission
    \item Fermi Gamma-ray Space Telescope
    \item Gamma-ray Bursts
    \item Dark Matter searches with X-ray and gamma-ray observations
    \item Multimessenger observations
    \item Pulsed signals
    \item Data Analysis Systems for Spectra
    \item AGN demography through the cosmic time
    \item Fundamental physics with neutron stars
    \item High mass X-ray binaries
    \item Galactic cosmic rays
    \item Pulsar wind nebulae
    \item Non-thermal processes and particle acceleration in supernova remnants
\end{itemize}

\end{document}